\documentclass[twocolumn,nofootinbib,superscriptaddress]{revtex4-1}
\usepackage{latexsym,epsfig,amssymb, amsmath,nicefrac}

\usepackage{color}
\usepackage[makeroom]{cancel}
  \usepackage{latexsym}
  \usepackage{epsf}
  \usepackage{amssymb}
  \usepackage{graphicx}
  \usepackage{amsmath}
  \usepackage{amsmath,amssymb,amsthm}
  \usepackage{verbatim}
  \usepackage{hyperref}

\def\tb0{\tilde{\beta}_0}
{\def\b0{\beta_0}

\def\bi{\begin{itemize}}
\def\ei{\end{itemize}}
\def\be{\begin{equation}}
\def\ee{\end{equation}}
\newcommand{\bea}{\begin{eqnarray}}
\newcommand{\eea}{\end{eqnarray}}

\def\Kahler{K\"{a}hler~}

\begin{document}

\begin{flushright}
CERN-TH-2016-184
\end{flushright}

%\vspace{1cm}

\title{General sGoldstino Inflation}

\author{Sergio Ferrara}
%\email{s.ferrara@cern.ch}
\affiliation{Theoretical Physics Department, CERN, CH-1211 Geneva 23, Switzerland}
\affiliation{INFN-Laboratori Nazionali di Frascati, Via Enrico Fermi 40, 00044 Frascati, Italy}
\affiliation{Department of Physics and Astronomy and Mani L.Bhaumik Institute for Theoretical Physics, U.C.L.A., Los Angeles, CA 90095-1547, USA}
\author{Diederik Roest}
%\email{d.roest@rug.nl}
\affiliation{Theoretical Physics Department, CERN, CH-1211 Geneva 23, Switzerland}
\affiliation{Van Swinderen Institute for Particle Physics and Gravity, University of Groningen, \\ Nijenborgh 4, 9747 AG Groningen, The Netherlands}

\begin{abstract}
We prove that all inflationary models, including those with dark energy after the end of inflation, can be embedded in minimal supergravity with a single chiral superfield. Moreover, the amount of supersymmetry breaking is independently tunable due to a degeneracy in the choice for the superpotential. The inflaton is a scalar partner of the Goldstino in this set-up. We illustrate our general procedure with two examples that are favoured by the Planck data.
\end{abstract}	

\maketitle

\section*{Introduction}

The framework of inflation is both theoretically appealing  and observationally succesful. The latest CMB observations of Planck 2015 provide a very accurate measurement of the spectral index of scalar perturbations: $n_s = 0.968 \pm0.006$ \cite{Planck}. In addition, constraints have been put on the remaining prediction of tensor perturbations: $ r = A_t / A_s < 0.07$ at $2\sigma$ \cite{KECK}. Despite this success, it remains imperative to embed inflation in a UV-complete theory of quantum gravity.

A first step in this direction is provided by supergravity. The embedding of inflation in such theories turns out to be surprisingly intricate, due to their non-trivial scalar potentials: these entail an interplay between the negative definite gravitino mass term and the positive definite supersymmetry (SUSY) breaking contributions. It took a decade and a half before the simplest model of quadratic inflation in supergravity was constructed \cite{Kawasaki}. This paper provided two improvements over previous constructions. 

First of all, it uses a shift-symmetric \Kahler potential $K = -\tfrac12 (\Phi - \bar \Phi)^2$ for the flat scalar manifold to avoid  the $\eta$-problem. With the conventional choice $K = \Phi \bar \Phi$, the scalar potential's overall factor exp$(K)$  gives order-one contributions to the second-slow roll parameter $\eta$ that ruin the slow-roll conditions \cite{Copeland}. Secondly, the construction employs an additional chiral multiplet $S$ that constitutes the sGoldstino, the scalar partner of the Goldstino field arising due to spontaneous SUSY breaking. This allows for an independent tuning of the gravitino mass and SUSY breaking, and thus decouples the two ingredients of the scalar potential.

Supergravity poses strong constraints on the (s)Goldstino fields: e.g.~the masses of this multiplet are set only by the SUSY breaking order parameter. The construction of \cite{Kawasaki} leaves the inflaton multiplet $\Phi$ free of such constraints. Indeed, it was realized a decade later that this construction allows for  arbitrary inflationary potentials \cite{Kallosh:2010, Rube}. The subsequent development of $\alpha$-attractors, providing an excellent agreement with the Planck results \cite{Planck}, builds on the same models while replacing the flat with a hyperbolic  \Kahler manifold instead \cite{attractors, Carrasco}.

The versatility of the constructions with the sGoldstino multiplet does not adress the question whether it is necessary to extend the field content in this manner. Models of sGoldstino inflation - with inflation taking place in the same direction in field space as SUSY breaking - have been investigated with varying results, see e.g.~\cite{Jimenez1, Jimenez2, Dine} and most recently \cite{Achucarro}. A common issue in these investigations is the interplay between inflation and SUSY breaking. The same issue was addressed in \cite{Palma1, Borghese} depending on the angle between the two directions of inflation and the sGoldstini (see also \cite{Palma2}). More recently, single-field realizations of $\alpha$-attractors were put forward \cite{Scalisi, Linde}.

In this letter we will prove a comparable versatility for sGoldstino inflation: all inflationary potentials can be realized in a single-superfield construction. The interplay between inflation and SUSY breaking therefore poses no constrains on the inflationary predictions  nor on the level of SUSY breaking, either during or after inflation. Moreover, we will demonstrate that one can independently introduce a cosmological constant in the final vacuum to describe the dark energy of the late Universe. The single-superfield framework, which can be thought of as arising after a supersymmetric decoupling of any other multiplets \cite{deAlwis, Sousa, Hardeman, Achucarro}, thus proves to be remarkably succesful in accomodating different physical phenomena.

\section*{\Kahler preliminaries}

The scalar dynamics of general supergravity models is specified in terms of two quantities. The real \Kahler potential $K=K(\Phi,\bar \Phi)$ specifies the \Kahler geometry of the scalar manifold in terms of the metric $K_{\Phi \bar \Phi}$, while the holomorphic superpotential $W=W(\Phi)$ (together with $K$) determines the scalar potential via
 \begin{align}
   V = e^K \left( - 3 W \bar W + K^{\Phi \bar \Phi} D_\Phi W D_{\bar \Phi} \bar W \right) \,,
 \end{align}
with the \Kahler metric $K^{\Phi \bar \Phi} \equiv K_{\Phi \bar \Phi}^{-1}$ and the \Kahler covariant derivative $D_\Phi W \equiv \partial_\Phi W + K_\Phi W$. The latter are covariant under the \Kahler transformations $W \rightarrow e^f W$ and $K \rightarrow K - f - \bar f$ with holomorphic $f(\Phi)$, which leave the scalar potential invariant. Indeed the entire supergravity theory can be written in terms of the \Kahler function $G = K + \log(W \bar W)$  \cite{Girardello} and is therefore manifestly invariant under this gauge transformation. The scalar potential reads
 \begin{align}
  V = e^G \left( -3 + G^{\Phi \bar \Phi} G_\Phi G_{\bar \Phi} \right) \,, \label{V-Kahler}
 \end{align}
in terms of this variable. Note that $G^{\Phi \bar \Phi} = K^{\Phi \bar \Phi}$ as the superpotential contributions drop out. 

Following \cite{Rube}, we will make the following assumptions about these potentials:
 \begin{itemize}
  \item
  The \Kahler potential will be invariant under $\Phi \rightarrow \bar \Phi$, i.e.~$K(\Phi, \bar \Phi) = K(\bar \Phi, \Phi)$, and hence will be an even function of the imaginary component of $\Phi$.
 \item
 The superpotential will be a real holomorphic function of $\Phi$, i.e.~$\overline{W(\Phi)} = W (\bar \Phi)$. 
 \end{itemize}
This implies that the \Kahler invariant function satisfies $G(\Phi, \bar \Phi) = G(\bar \Phi ,\Phi)$. These properties guarantee that the truncation to $\Phi = \bar \Phi$, specifying the trajectory along which inflation will take place, will be a consistent one: the field equation for the imaginary component of $\Phi$ is satisfied along the trajectory $\Phi = \bar \Phi$ for arbitrary choices of \Kahler and superpotentials satisfying the above criteria\footnote{The constructions of \cite{Ketov1, Ketov2} instead use the imaginary direction as inflaton and require higher-order terms to achieve stability.}. Moreover, one can employ the \Kahler transformation to simplify the \Kahler potential:
 \begin{itemize}
 \item 
  The \Kahler potential vanishes, $K=0$, along the inflationary trajectory. This consistutes a shift symmetry of the inflaton field and implies that $K_\Phi = 0$ along $\Phi = \bar \Phi$.
 \end{itemize}
Note that this can be achieved for arbitrary \Kahler potentials, and fully fixes the \Kahler frame. It should be stressed that this is not an additional physical requirement on the theory and only amounts to a convenient gauge fixing. Examples that will be relevant are
 \begin{align}
   K_{\rm fl} = -\tfrac12 (\Phi - \bar \Phi)^2 \,, \quad K_{\rm hy} = -3 \alpha \log \left( \frac{T + \bar T}{2 |T|} \right) \,, \label{Kahler}
 \end{align}
describing a flat and hyperbolic manifold in terms of coordinates $\Phi$ and $T$, respectively.

\section*{Superpotential flows}

\begin{figure}[t!]
%	\vspace{0.4cm}
	\begin{center}
		\includegraphics[width=3.5cm]{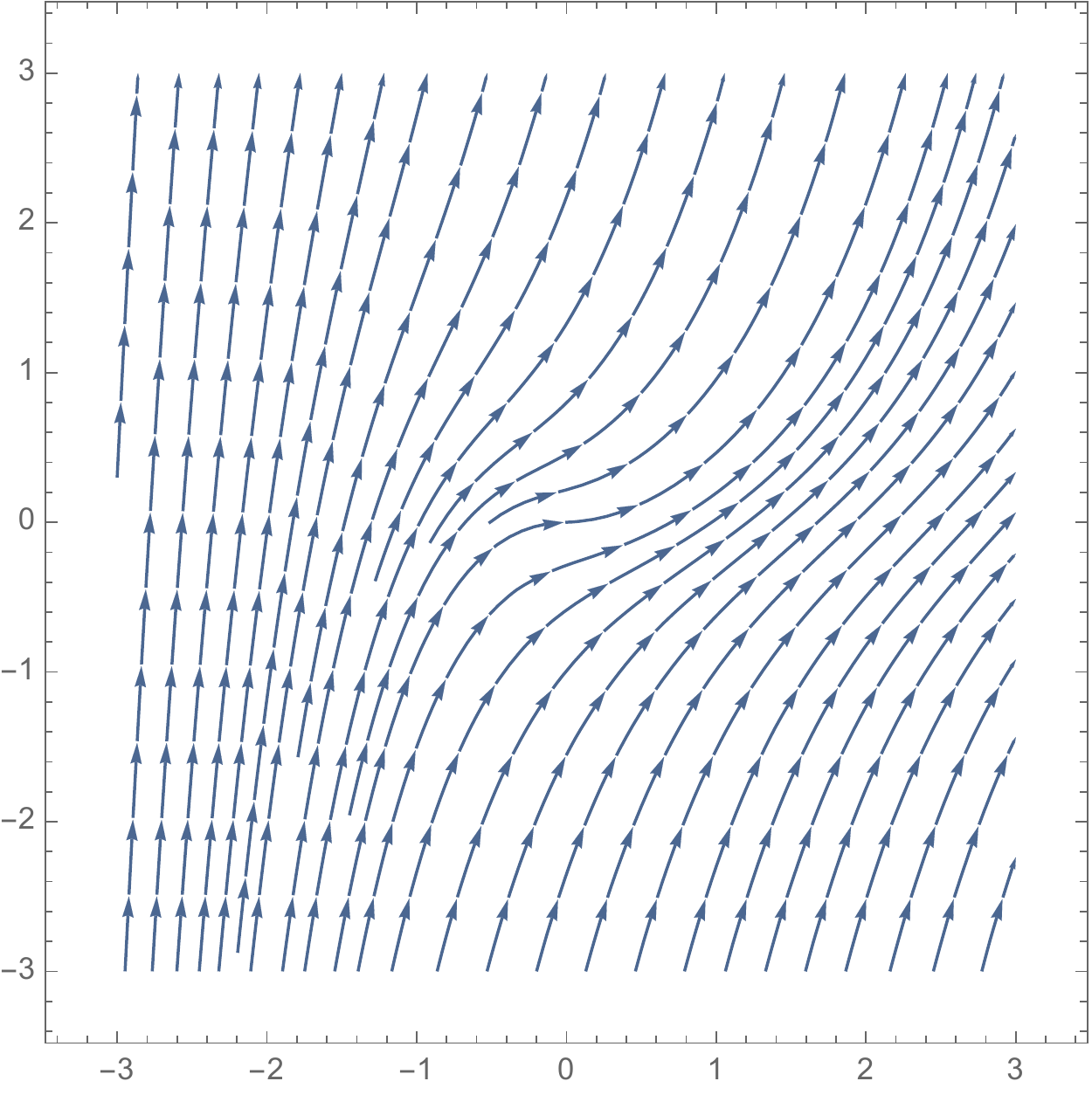}
		\includegraphics[width=5cm]{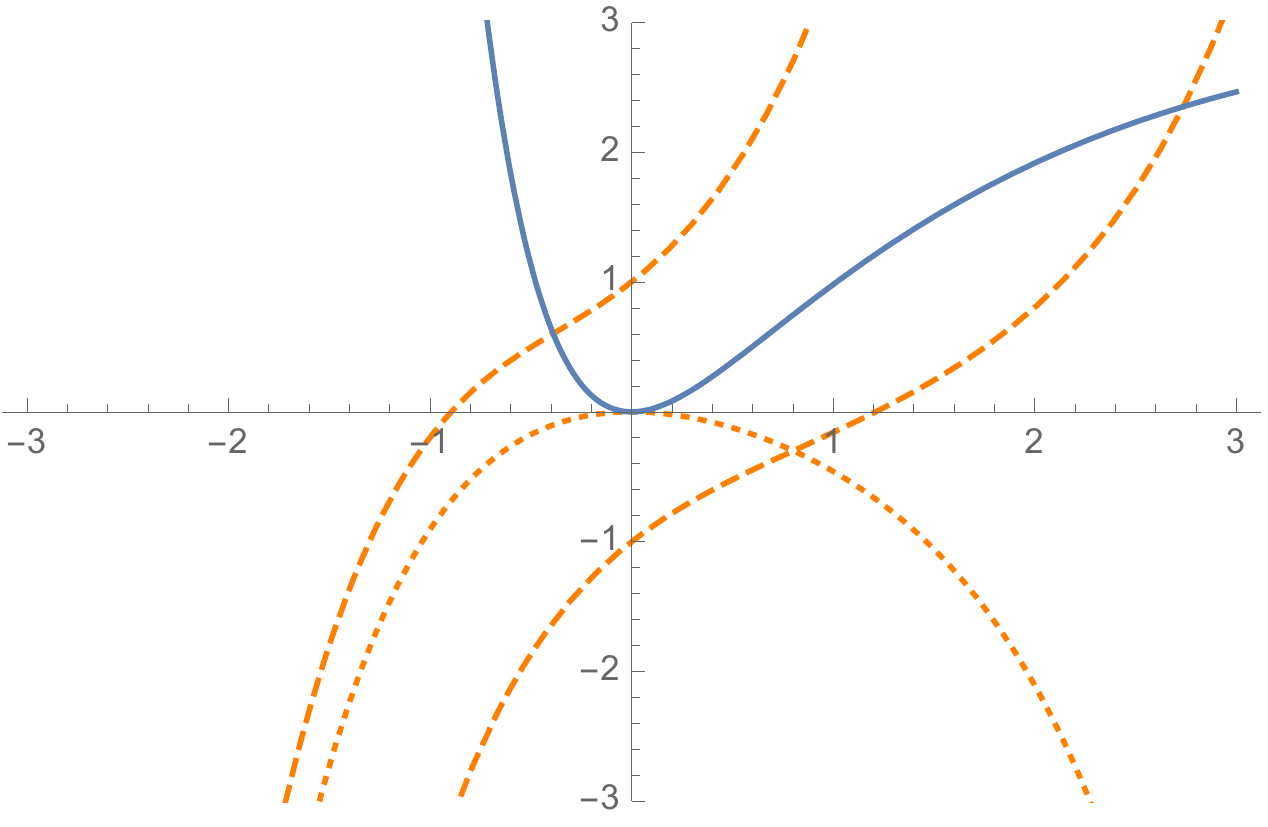}
		\caption{\it Left: the flow \eqref{flow} in $(\varphi,W(\varphi))$ for a given scalar potential with a Minkowski minimum. Right: three different superpotentials in orange (the SUSY one dotted and the non-SUSY ones dashed) for the same flow with the resulting scalar potential in solid blue.}
		\label{fig:flow}
	\end{center}
	\vspace{-0.8cm}
\end{figure}

Along the inflationary trajectory, the scalar manifold is one-dimensional and hence can always be brought to canonical form by means of a field redefinition $\Phi = \Phi(\varphi)$. For the examples of \Kahler geometries above one finds
 \begin{align}
  \Phi = \varphi / \sqrt{2} \,, \quad T = e^{-\sqrt{2/(3 \alpha)} \varphi} \,.
 \end{align}
In terms of the canonically normalized inflaton field $\varphi$, the superpotential $W=W(\Phi(\varphi))$ therefore satisfies
 \begin{align} \label{flow}
  \frac{d W}{d \varphi} = \pm \sqrt{\frac{3W^2}{2} + \frac{V(\varphi)}{2}} \,. 
 \end{align}
Given a specific scalar potential, this is a non-autonomous first-order differential equation for $W(\varphi)$. We will assume $V$ to be non-negative everywhere with a minimum that can be either Minkowski or De Sitter. Without loss of generality one can take it at $\varphi=0$. 
Then the above equation always has a one-parameter family of solutions: given a value $W(\varphi_0)$ at some point, one can follow the flow defined by \eqref{flow} and illustrated in figure \ref{fig:flow}. The flow is always well-defined since $V$ is non-negative. The different initial conditions $W(\varphi_0)$ result in different superpotential yielding the same scalar potential. This can be seen as the analogon of 'fake supergravities' \cite{Freedman} describing domain wall solutions in AdS dual to RG-flows, see also \cite{Kiritsis1, Kiritsis2}.

A special feature occurs in the case of a Minkowski minimum: the flow becomes horizontal at $W(\varphi=0) = 0$. Note that this corresponds to the choice of a supersymmetry preserving Minkowski minimum; all other choices would have the same Minkowski minimum with broken supersymmetry. The horizontal flow at $V=W=0$ signals that one has to take the other branch of solutions, with opposite sign in \eqref{flow}, to continue in $\varphi$-space (using the same signs on both sides of $\varphi$ would result in a discontinuous second derivative of the superpotential). We will refer to this as the SUSY superpotential. This is illustrated this for a specific $V(\varphi)$ in figure \ref{fig:flow}. 

We would like to stress that one can embed {\it any} non-negative scalar potential in this way in a single-superfield, allowing for a description of inflation (be it of the chaotic, hilltop or plateau type) as well as dark energy. Moreover, the one-parameter ambiguity in the scalarpotential for a given $V(\varphi)$ allows one to introduce SUSY breaking independently. The counterpart to this simplicity is a non-autonomous differential equation. While this always has a solution and allows for a simple numerical treatment, it can be hard to find a closed expression for $W$. 

Its asymptotic behaviour, however, follows from the observation that the superpotential at large $\varphi$ increases at least as fast as exp$(\sqrt{3/2} \varphi)$. Viable inflationary potentials are much flatter, and hence $V$ will be a small correction to $W^2$ in the flow. Note that this implies that the gravitino mass and the scale of SUSY breaking exceed the Hubble scale by far. In this regime, the superpotential can be expanded as
 \begin{align}
  W = \pm W_0 e^{\sqrt{3/2} \varphi} \left( 1 - e^{- \sqrt{6} \varphi} V(\varphi) + \ldots \right), \label{approximation}
 \end{align}
in terms of exp$(- \sqrt{6} \varphi) V$. Such superpotentials were put forward in combination with a flat \cite{GL, First} and hyperbolic \cite{Scalisi, Linde} \Kahler geometry to describe plateau inflation, while the present context is fully general: we have not made any assumptions on the scalar nor the \Kahler potential. Moreover, a similar expansion holds at large $-\varphi$. 

The aformentioned difference between SUSY and non-SUSY superpotentials translates into the asymptotic signs
  \begin{align}
    {\rm SUSY:~} & W = W_0 (  e^{\pm \sqrt{3/2} \varphi} + \ldots ) \,, \quad x \rightarrow \pm \infty \,, \notag \\
    {\rm non-:~} & W = \pm W_0 (  e^{\pm \sqrt{3/2} \varphi} + \ldots ) \,, \quad x \rightarrow \pm \infty \,. \label{asymptotics}
  \end{align}
of which the SUSY one touches $W=0$ at the minimum\footnote{A direct corollorary is that any superpotential that has the asymptotics \eqref{asymptotics}  while it is non-zero at the minimum $\varphi=0$ has an AdS instead of Minkowski vacuum.} $\varphi=0$. In contrast, generic, non-SUSY superpotentials have opposite signs in both asymptotic regions and cross through $W=0$ at a different point $\varphi \neq 0$. Prototypical examples are $\cosh(\sqrt{3/2} \varphi)$ and $\sinh(\sqrt{3/2} \varphi)$, respectively.

The special role of the SUSY superpotential resolves an apparent paradox with \cite{Wrase}, which shows that infinitesimal transformations of the superpotential cannot deform a supersymmetric Minkowski minimum into a non-SUSY one. Due to the sign flip of the SUSY solution, its superpotential indeed has a non-continuous difference from the non-SUSY solutions.

We would also like to mention that our models have holomorphic superpotentials depending on a complex scalar field $\Phi$. The reason that these can be converted into real functions $W(\varphi)$    is the assumption that $W(\Phi)$ is a real holomorphic function. This ensures that the truncation to a real variable $\varphi$ is consistent and unambiguous. Vice versa, solutions to the flow equation \eqref{flow} allow one to reconstruct the full holomorphic superpotential.

\section*{Stability issues}

We have thus identified a large degeneracy in the choice of both the \Kahler and the superpotential to generate a specific inflationary model along the real line. However, these models will have different properties away from $\Phi = \bar \Phi$. An important aspect of this concerns the stability of the truncation to the inflaton field: what is the mass of the orthogonal direction, the sinflaton\footnote{If the inflaton is embedded in a massive vector multiplets with a D-term potential, the issue of stability does not arise for any \Kahler manifold. A particular example is the Starobinsky model from $R+R^2$ supergravity in the new minimal formulation \cite{Porrati}.}? 

From the general scalar potential \eqref{V-Kahler} it follows that the average of the scalar masses is given by
 \begin{align}
  m^2 = & G^{\Phi \bar \Phi} V_{\Phi \bar \Phi} = e^G \Big( -2 + G^{\Phi \bar \Phi}  G_{\Phi}  G_{\bar \Phi} R + \notag \\
  & + (G^{\Phi \bar \Phi})^2 (G_\Phi^2 + D_\Phi G_\Phi) (G_{\bar \Phi}^2 + D_{\bar \Phi} G_{\bar \Phi}) \Big) \,, \label{mass-original}
 \end{align}
where $R = G^{\Phi \bar \Phi} R_{\Phi \bar \Phi} = - G^{\Phi \bar \Phi} \partial_\Phi \partial_{\bar \Phi} \log(G_{\Phi \bar \Phi})$ is the curvature of the \Kahler manifold, which is given by $-2/(3\alpha)$ for the hyperbolic case and vanishes for the flat case. Note that, in the SUSY limit with $G_\Phi = 0$, the above formula relates the spin-0 and spin-1/2 masses in an Anti-de Sitter background. In the non-SUSY case, it can be rewritten as
 \begin{align}
  m^2 = & e^G \left( 2 + G^{\Phi \bar \Phi} G_\Phi G_{\bar \Phi} R \right) + \notag \\
   & + \frac{2 V_\Phi G_{\bar \Phi} + 2 V_{\bar \Phi} G_{\Phi} + e^{-G} V_\Phi V_{\bar \Phi} }{G_{\Phi} G_{\bar \Phi}} \,,
 \end{align}
which is a generalization of the mass supertrace relations of \cite{Cremmer, Grisaru} that is valid for any $V$ and any $V_\Phi$, and therefore also during inflation. In the present set-up, in terms of the real variable $\varphi$ this reduces to
 \begin{align}
  m^2 = 2 W^2 + 2 W_\varphi^2 R + \frac{V_\varphi^2}{4 W_\varphi^2} + \frac{2 W V_\varphi}{W_\varphi} \,. \label{mass}
\end{align}
Away from a critical point, the latter two terms are corrections proportional to $V'$; however, during slow-roll inflation these will be suppressed compared to $V$. The same applies to the mass of the inflaton field, which is set by $V''$. In turn, $V$ and hence the Hubble scale is asymptotically much smaller than $W^2$. The first two terms of this expression above therefore set the mass of the orthogonal sinflaton direction during slow-roll inflation:
 \begin{itemize}
 \item For $R>-2/3$ or $\alpha>1$ the model is stable, with the sinflaton mass$^2$ increasing with the gravitino mass. This includes the flat \Kahler case.
 \item For $R<-2/3$ or $\alpha<1$ the model is unstable, with the sinflaton mass$^2$  decreasing with the gravitino mass.
 \item For $R=-2/3$ or $\alpha=1$ the model is unstable, with the sinflaton mass$^2$ decreasing with the Hubble scale. An example is $f(\mathcal{R})$ supergravity extensions with an F-term action in terms of the chiral scalar curvature multiplet $\mathcal{R}$ \cite{Ketov, FKV}.

 \end{itemize}
Note that $\alpha=1$ is exactly the dividing line between both asymptotic behaviours where the leading term in \eqref{mass} vanishes. However, for $\alpha \leq 1$ one can include higher-order stabilizer terms \cite{Linde, Carrasco} in order to achieve asymptotic stability.

Apart from the model-independent analysis during slow-roll inflation, the check of stability over the entire inflationary range is model-dependent and needs to be checked for specific cases. Once one ends up in a SUSY Minkowski minimum, the original mass formula \eqref{mass-original} implies that the masses of both scalar components (which are equal in this vacuum) are set by the spin-1/2 mass terms and are independent of the \Kahler curvature.

As a sideremark, in the rigid limit (where there is no \Kahler connection), the mass formula takes the particularly simple form
 \begin{align}
  m^2 = VR + \frac{K^{\Phi \bar \Phi} V_\Phi V_{\bar \Phi}}{V} \,,
 \end{align}
where $V = K^{\Phi \bar \Phi} W_\Phi \bar W_{\bar \Phi}$. This implies that a massive sGoldstino can only be stable with a non-vanishing and positive curvature; $R=0$ is the dividing line between (in-)stability in the rigid limit. An example is the linear realization of the Volkov-Akulov model, as shown by \cite{Komargodski}. In contrast to their set-up, where both scalar components have comparable masses and hence can be made very massive, we are employing a flat \Kahler potential to generate a very light inflaton field. Therefore it is impossible to take a non-linear limit of the present model.
 
\section*{Example I: Starobinsky inflation}

We will illustrate the general procedure in two concrete examples. The first example concerns the Starobinsky model of inflation \cite{Starobinsky}. Originally formulated as an $R+R^2$ theory, it can be rephrased in terms of a canonical scalar field with scalar potential
 \begin{align}
  V = 3 H^2 \left(1 - e^{-\sqrt{2/3} \varphi} \right)^2 \,. \label{Staro}
 \end{align}
We will set $H=1$ in all subsequent plots. 
At large positive field values, the potential asymptotes to a plateau with an exponentially suppressed fall-off. As a result, the Starobinsky model shares its inflationary predictions $n_s = 0.97$ and $r=0.003$ at $N=60$ e-folds with a large class of inflationary models with other origins \cite{Higgs, conformal, universal}. It was demonstrated in \cite{Scalisi, Linde} that such plateau models can be described by a superpotential of the form\footnote{A superpotential that is dominated by a single monomial cannot generate a De Sitter plateau \cite{Ellis}.}
  \begin{align}
   W = W_0 (T^{-3/2} - T^{3/2} f(T) ) \,,
 \end{align}
in terms of hyperbolic coordinates $T$ with \Kahler potential \eqref{Kahler}, where the profile function $f(T)$ has a regular Taylor expansion around $T=0$. In order to obtain the full Starobinsky potential, it turns out that one must choose an inflationary profile that also has a regular Taylor expansion around $T = \infty$.

As discussed before, although its existence can be proven, obtaining an exact expression for the underlying superpotential can be very hard. In this case, however, we can obtain a very good approximation by making an Ansatz for the profile function given by a Pade approximant,
 \begin{align}
  f(T) = \frac{a + b T}{1 + c T} \,,
 \end{align}
compatible with both asymptotic limits $T \rightarrow 0$ and $T\rightarrow \infty$. Requiring the resulting scalar potential to have a Minkowski vacuum at $\varphi=0$ yields two constraints on the three parameters of this Ansatz. The remaining parameter interpolates between different approximations of the Starobinsky potential \eqref{Staro}. For concreteness, we will take a specific element with coefficients $(a,b,c) = (7,-5,1)$. This particular superpotential was actually plotted as the dotted orange line in figure~\ref{fig:flow}.

\begin{figure}[t!]
%	\vspace{0.4cm}
	\begin{center}
		\includegraphics[width=4.25cm]{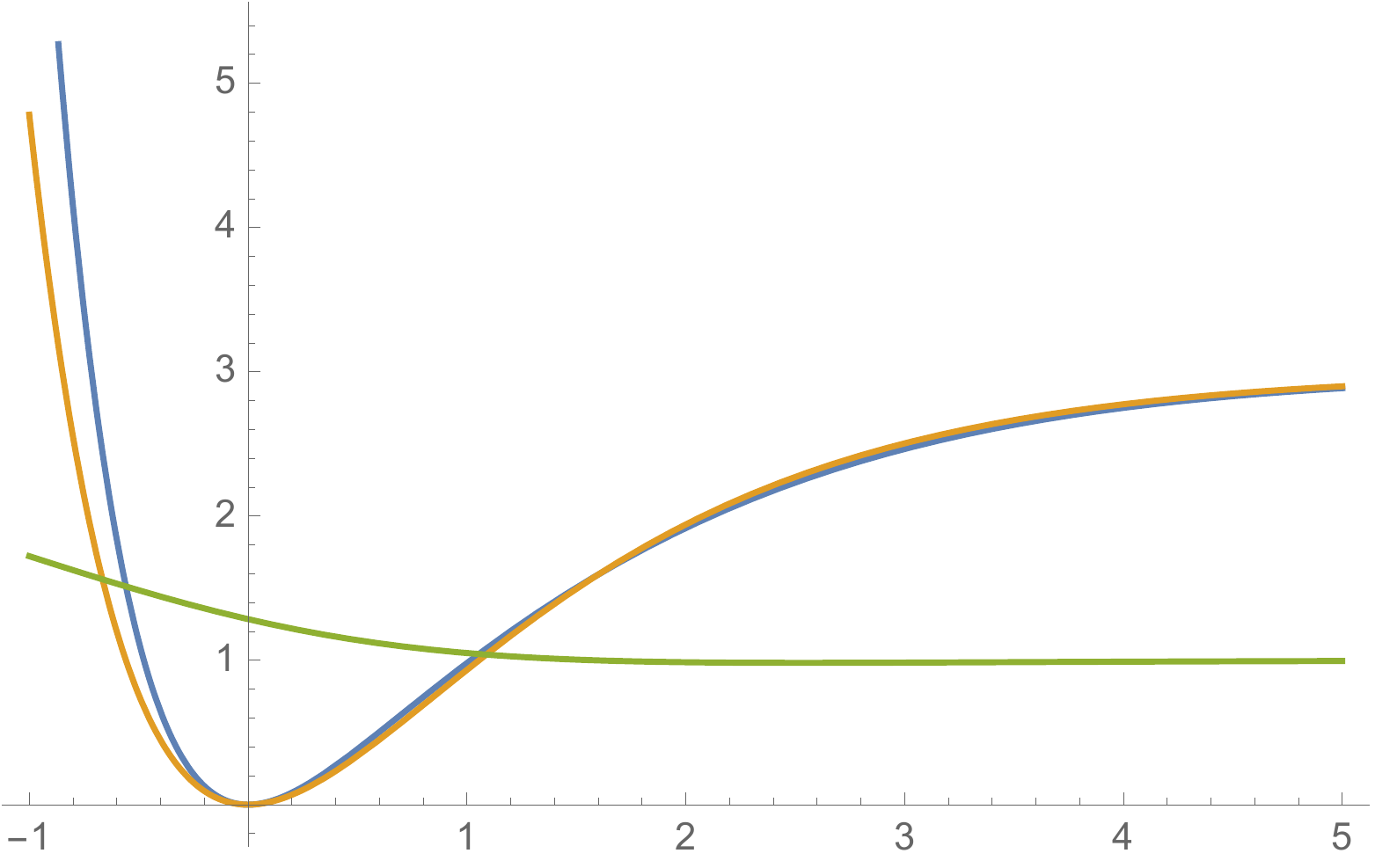}
		\includegraphics[width=4.25cm]{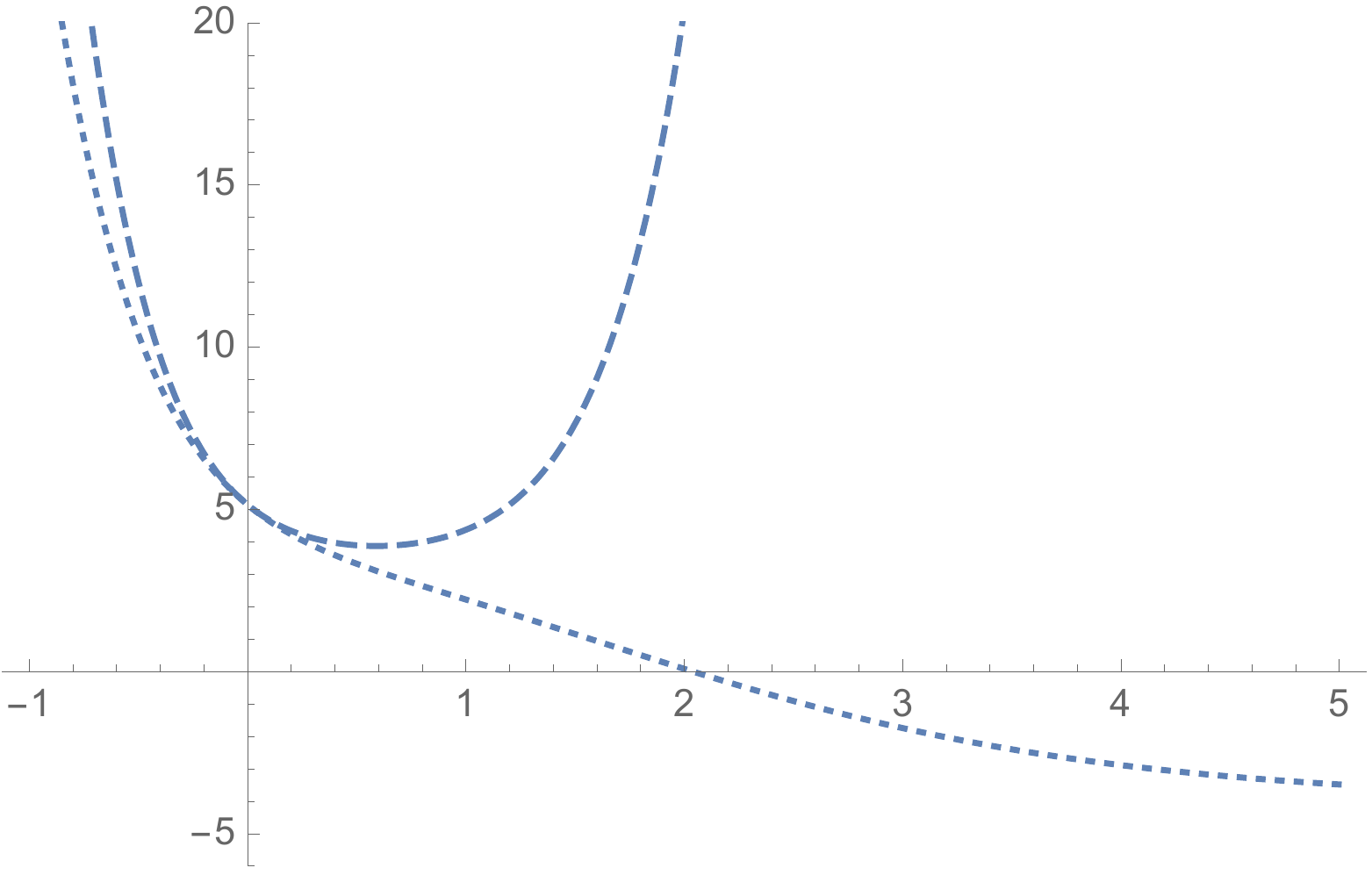}
		\caption{\it Left: the scalar potential of Starobinsky inflation (orange) and its approximation in terms of a Pade superpotential (blue). We also indicate the ratio between the two (green). Right: the mass $m^2$ for the imaginary inflaton partner in the case of a Pade superpotential and a hyperbolic (dashed) and flat (dotted) \Kahler potential.}
		\label{fig:Staro}
	\end{center}
	\vspace{-0.8cm}
\end{figure}

The resulting scalar potential is equal to Starobinsky inflation up to a field-dependent factor $V = d(\phi) V_{\rm St}$ with
 \begin{align}
  d(\phi) = & \frac{1}{109 \cosh^4(\phi/\sqrt{6})} \big( 43 \cosh(\sqrt{\frac83} \phi) + 74 \cosh(\sqrt{\frac23} \phi)  \notag \\ 
& +  27  - 22 \sinh(\sqrt{\frac23} \phi) - 13 \sinh(\sqrt{\frac83} \phi) \big) \,.
 \end{align}
Note that $d(\phi)$ is a smooth function that never vanishes and interpolates between the values $1$ and $179/109$ in an almost monotonic manner. From the moment of CMB all the way up to the end of inflation, it only changes a few percent, and hence the inflationary predictions will be virtually identical to those of Starobinsky (Fig.~\ref{fig:Staro}). This demonstrates that one can obtain arbitrarily accurate approximations to this particular inflationary model. Moreover, the general argument put forward in this letter guarantees that there is a superpotential that exactly reproduces the scalar potential \eqref{Staro}.

Turning to the stability of the orthogonal direction, the general discussion implies that plateau models with $\alpha=1$ will have an instability when embedded in a hyperbolic \Kahler geometry \eqref{Kahler}. This can be remedied by going to a flat \Kahler metric. The relation to $\varphi$ on the real line implies that one must use the coordinate replacement
 \begin{align}
   T = e^{- 2 \Phi / \sqrt{3 \alpha}} \,. \label{coordinates}
 \end{align}
with $\alpha=1$ to obtain the corresponding superpotential in flat coordinates:
 \begin{align}
 \frac{W}{W_0} = \sinh( \frac{\Phi}{\sqrt{3}} ) \tanh( \frac{\Phi}{\sqrt{3}} ) \left( 1 + 6 \cosh( \frac{2\Phi}{\sqrt{3}} )  - 4 \sinh( \frac{2\Phi}{\sqrt{3}} ) \right) \,.
 \end{align}
Note that the superpotential encounters a singularity when the imaginary component of $\Phi$ equals $\pm \sqrt{3} \pi /2$; indeed the scalar potential becomes infinite at this line. We therefore have to restrict to a horizontal strip of the flat geometry. 

As follows from the general discussion, the imaginary direction will be stable on the plateau at large $\Phi = \bar \Phi$ for the flat model. One can check that the same holds along the entire inflationary line (Fig.~\ref{fig:Staro}). Note that the sinflaton masses are equal in the SUSY Minkowski minimum for the hyperbolic and flat cases, and asymptote to the negative Hubble scale or a positive gravitino mass during inflation, respectively. All these aspects follow from the general discussion on the stability issues.  

\section*{Example II: Goncharov-Linde inflation}

An inflationary model with a simpler superpotential, allowing for a nice illustration of our general considerations, is actually provided by the first supergravity model of inflation by Goncharov and Linde (GL) \cite{GL, First}. It has a flat \Kahler geometry and a superpotential that reads
 \begin{align}
  W = \frac{1}{\sqrt{3}} \sinh(\sqrt{3} \Phi) \tanh (\sqrt{3} \Phi) \,,
 \end{align}
where one has to restrict the imaginary component of $\Phi$ to the domain between $\pm \pi /(2 \sqrt{3})$. Along the middle of this strip, $\Phi = \bar \Phi$, the resulting scalar potential reads
 \begin{align}
 V = \tanh^2(\sqrt{3/2} \varphi) \big( 4 - \tanh^2 (\sqrt{3/2} \varphi) \big) \,.
 \end{align}
Again this has a plateau, however, with a different exponential fall-off than the Starobinsky model\footnote{There appears to be an interesting relation between the two models: they both follow from a similar Ansatz. In terms of curved coordinates \eqref{coordinates}, this class is given by the general Ansatz of \cite{Scalisi} with the specific Pade approximant choice
 \begin{align}
  f(T) = 1 + 6 \sqrt{\alpha} \frac{1-T}{1+T} \,,
 \end{align}
which was considered for different reasons in \cite{Linde}.}.
It therefore belongs to the same class of models as $\alpha$-attractors with $\alpha=1/9$.

\begin{figure}[t!]
%	\vspace{0.4cm}
	\begin{center}
		\includegraphics[width=4.25cm]{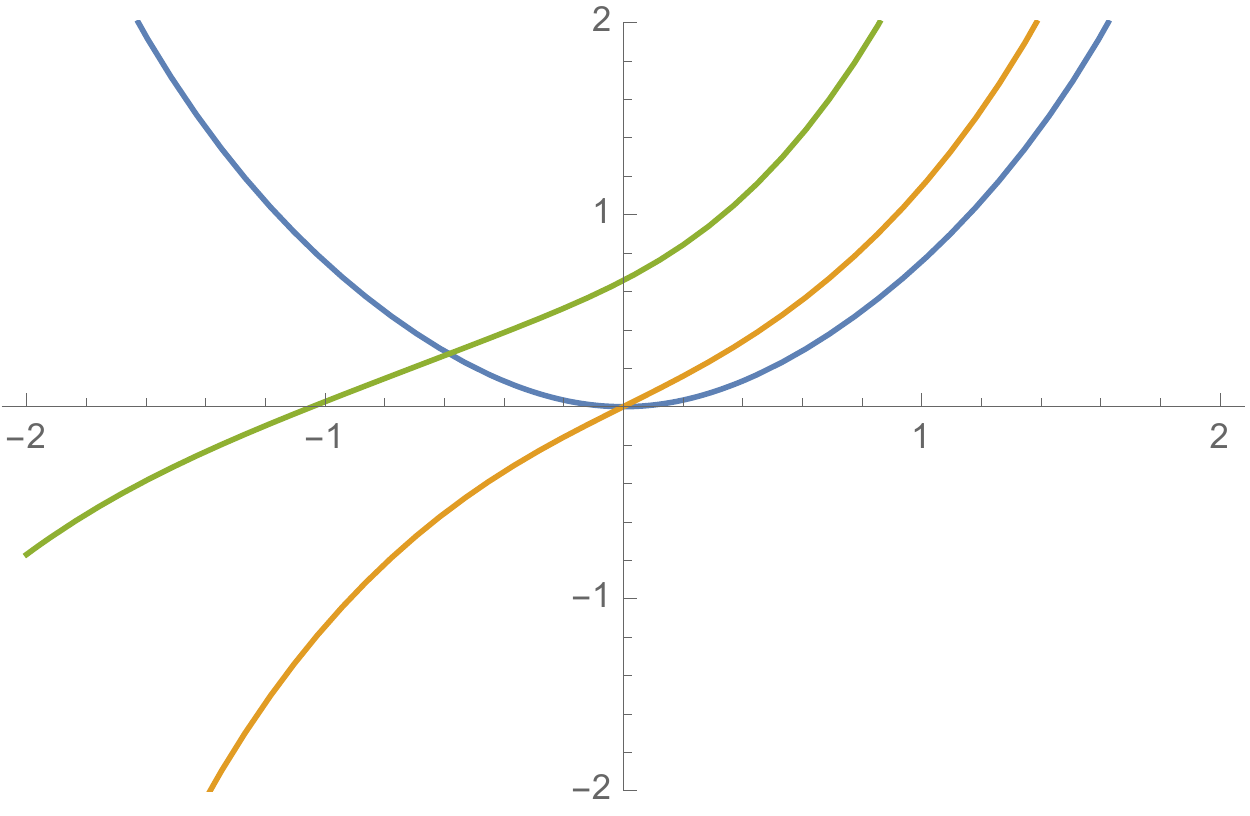}
		\includegraphics[width=4.25cm]{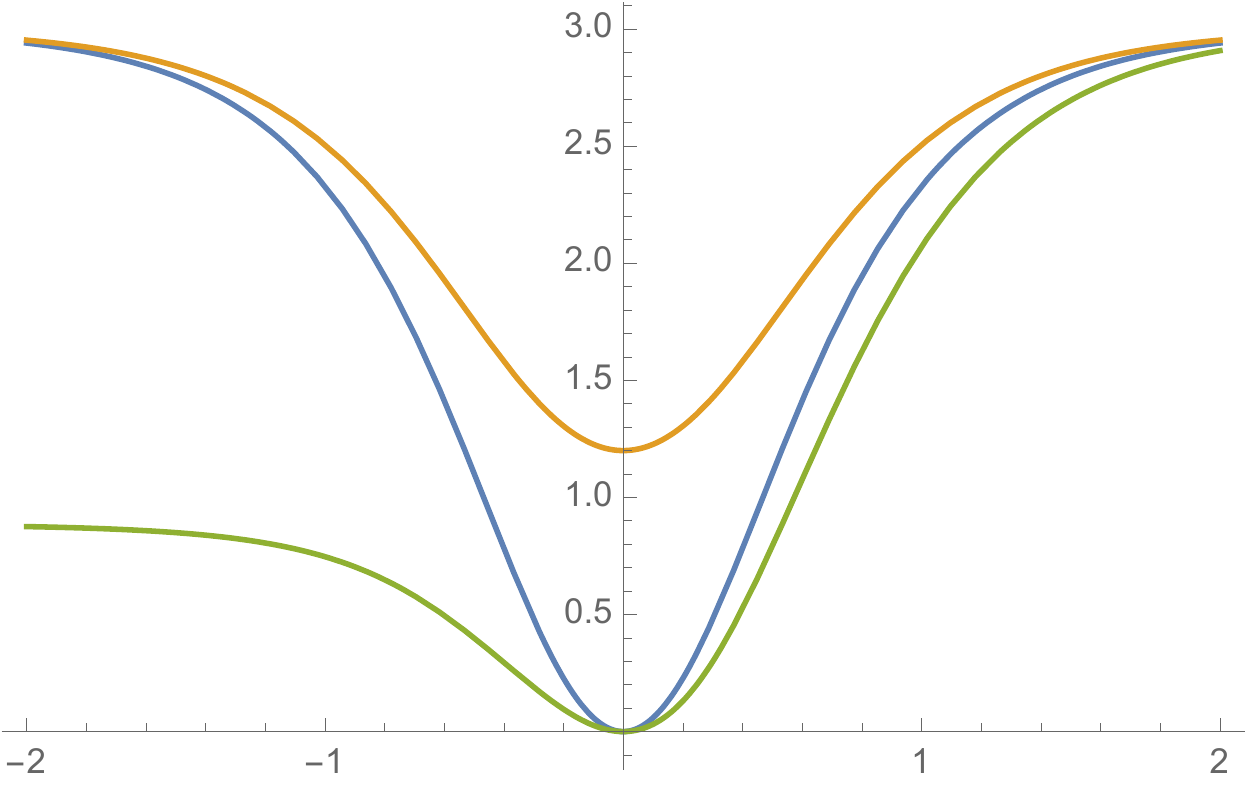}
		\caption{\it The superpotentials (left) and scalar potentials (right) for the generalized GL-model with parameter choices $(\epsilon, \delta) = (0,0), (1,0), (1,\sqrt{2/3})$ in blue, orange and green, respectively.}
		\label{fig:valley}
	\end{center}
	\vspace{-0.8cm}
\end{figure}

The original GL model has a Minkowski minimum with unbroken SUSY. We will introduce two independent deformations of this model in order to describe SUSY breaking and a positive cosmological constant. The generalized superpotential is
 \begin{align}
  \frac{W}{W_0} =  \sqrt{\sinh(\sqrt{3} \Phi)^2 + \tfrac23 \epsilon^2} \tanh (\sqrt{3} \Phi) + \delta \cosh (\sqrt{3} \Phi) \,. \label{generalized-GL}
 \end{align}
The resulting scalar potential still has a minimum at $\varphi = 0$, at which
 \begin{align}
   \frac{V}{W_0^2} = 2 \epsilon^2 - 3 \delta^2 \,, \quad
   \frac{W}{W_0} = \delta \,, \quad
   \frac{\partial_\varphi W}{W_0} = \epsilon \,.
 \end{align}
Non-negativity of the scalar potential requires $2 \epsilon^2 - 3 \delta^2$ to be zero or positive. The first case leads to a Minkowski vacuum, while in the second case the value of this combination determines the cosmological constant. Independently of this, one can introduce SUSY breaking by means of the parameter $\epsilon$. The different parameter choices are illustrated in Fig.~\ref{fig:valley}.

\section*{Discussion}

In this letter we have outlined a general framework to realize  arbitrary scalar potentials with a Minkowski or De Sitter vacuum as sGoldstino inflation. Moroever, the one-parameter degeneracy in the superpotential corresponds to the freedom to incorporate SUSY breaking. Perhaps the clearest illustration is provided by the superpotential \eqref{generalized-GL} with separate parameters for SUSY breaking and the cosmological constant. In addition, we have discussed an accurate approximation of Starobinsky inflation.

A general feature of our models is that the scale of SUSY breaking, set by $W_\varphi$, is much larger than the inflationary Hubble scale, set by $V$. One might worry that this exceeds the Planck scale. However, due to the hierarchy between the Hubble and the Planck scale this does not happen until $\varphi \gtrsim \sqrt{2/3} \log(M_{\rm Pl} / H) \sim 10$ in Planck units. Therefore this allows for the observationally preferred models of plateau inflation. Similarly, our general construction employs a \Kahler frame with a shift symmetry along the inflationary direction, which is however fully broken by the superpotential. This construction therefore requires a different mechanism to protect itself from quantum corrections with large implications for inflation.

\section*{Acknowledgements}

We would like to thank Renata Kallosh, Andrei Linde and Augusto Sagnotti for discussions and for collaboration on related work. SF is supported in part by INFN-CSN4-GSS.

%\newpage

\bibliography{single-superfield}
\bibliographystyle{JHEP}

\end{document}